\documentstyle[preprint,pre,aps]{revtex}
\draft
\begin{document}
\date{\today}
\title{PHASE ORDERING OF \\
2D XY SYSTEMS BELOW $T_{\text{KT}}$}
\author{A. D. Rutenberg and A. J. Bray}
\address{Theoretical Physics Group \\
Department of Physics and Astronomy \\
The University of Manchester, M13 9PL, UK\\
cond-mat/9411093}
\maketitle

\begin{abstract}
We consider quenches in non-conserved two-dimensional XY systems
between any two temperatures below the Kosterlitz-Thouless transition.
The evolving systems are defect free at coarse-grained scales,
and can be exactly treated.  Correlations
scale with a characteristic length $L(t) \propto t^{1/2}$ at late
times. The autocorrelation decay exponent,
$\bar{\lambda} = (\eta_i+\eta_f)/2$,
depends on both the initial and the final state of the quench
through the respective decay exponents of equilibrium
correlations, $C_{EQ}(r) \sim r^{-\eta}$.  We also discuss
time-dependent quenches.
\end{abstract}

Phase-ordering kinetics is the study of the non-equilibrium process of
equilibration after a rapid change of system parameters such as
temperature or pressure \cite{Bray94b}.  Typically, the system is
quenched from a
high-temperature disordered phase to a low-temperature ordered phase.
The problem is challenging, because there are
degenerate ground states competing to select the ordered phase, and
because the evolution of the order-parameter is typically determined
by a non-linear
partial differential equation. Indeed, there have been only a handful
of exact solutions in phase-ordering systems (see \cite{Bray94b}):
non-conserved  scalar and XY [$O(2)$] systems in one
dimension, and both non-conserved and conserved spherical models in
general dimensions.  Typically, phase-ordering systems have singular
topological defects (such as domain walls, vortex lines, or point
defects) seeded from the disordered initial conditions.
While the structure of singular defects can be used to extract
the growth laws of characteristic scales through Energy-Scaling
arguments \cite{Bray94a}, their singular nature makes exact solutions
 of correlation functions difficult.  In fact, the exact solutions
mentioned above do not involve singular defects: one-dimensional XY
systems have only non-singular
topological textures \cite{Rutenberg94}, while spherical systems have no
topological defects. It is natural, therefore, to consider other systems
without topological defects.

In this paper, we consider the coarse-grained two-dimensional (2D)
XY model below the Kosterlitz-Thouless
transition temperature $T_{KT}$, where there are no free vortices
\cite{Kosterlitz73}.  For non-conserved dynamics we
exactly determine two-point correlations after a quench between any two
temperatures at or below $T_{KT}$.
We find scaling solutions characterized by a single time-dependent
length  $L(t) \sim t^{1/2}$, without the logarithmic factor
[$L\sim (t/\ln{t})^{1/2}$] associated with quenches to states
with free vortices (i.e.\ from above $T_{KT}$) \cite{Bray94a,Yurke93}.
We also measure autocorrelations, for which the decay of the overlap
with the initial conditions is characterized by an exponent
$\bar{\lambda}$, through $A(t) \sim L^{-\bar{\lambda}}$
\cite{Fisher88,Majumdar94,Pargellis94}.
For quenches to $T=0$ we find $\bar{\lambda} = \eta_i/2$, where $\eta_i$
characterizes the initial asymptotic spatial correlations through
$C(r) \sim r^{-\eta_i}$.  This agrees with predictions of Bray
{\em et al.} for quenches from a critical point to zero temperature
\cite{Bray91}.  For quenches between arbitrary temperatures below
$T_{KT}$ we find $\bar{\lambda} = (\eta_i + \eta_f)/2$, which depends
on the
initial {\em and} final states of the quench [$\eta_f$ characterizes the
asymptotic {\em equilibrium} correlations at the final state through
$C_{EQ}(r)\sim r^{-\eta_f}$].   We also discuss quenches with arbitrary
temperature histories, and show that asymptotic correlations and
autocorrelations are independent of the early temperature history.

We consider over-damped, non-conserved, dissipative, ``model A''
dynamics with an equation of motion
\begin{equation}
\label{EQN:SPINDYN}
\partial_t \vec{\phi}=
 - \Gamma \delta H/\delta \vec{\phi} + \vec{\xi}({\bf x},t),
\end{equation}
where $\vec{\phi}({\bf x})$ is a two-component order-parameter.
The energy-functional,
\begin{equation}
\label{EQN:HAMILTONIAN}
	H[\vec{\phi}] = \int d^d x\,\left[ \frac{1}{2}
	(\nabla \vec{\phi})^2 + V(\vec{\phi})
	\right]\ ,
\end{equation}
has a potential with a symmetric global minimum at $|\vec{\phi}|=1$,
such as $V(\vec{\phi})= V_0 ( \vec{\phi}^2 -1)^2$. The
thermal noise, $\vec{\xi}= (\xi_1,\xi_2)$, is
Gaussian distributed with zero mean, with correlations determined
by the fluctuation dissipation theorem:
$\left< \xi_i({\bf x},t) \xi_j({\bf x}',t') \right>
= 2 \Gamma k_B T \delta^2({\bf x}-{\bf x}') \delta(t-t') \delta_{ij}$.
Below the Kosterlitz-Thouless transition
for the 2D XY model \cite{Kosterlitz73} any
vortices present will be bound in oppositely charged pairs. We only
consider
correlations at distances much larger than the characteristic pair size,
so that we need only treat the renormalized
spin-waves of the system. Effectively we
work on the line of zero-fugacity renormalization-group fixed points
with $0 \leq T \leq T_{KT}$.
In the limit $V_0 \rightarrow \infty$ the field, coarse-grained beyond
the pair scale, has unit magnitude.
We change to phase variables $\vec{\phi} = e^{i \theta}$, and the
energy-functional takes the form
\begin{equation}
\label{EQN:SWHAM}
 	H = \frac{\rho_s}{2} \int ( {\bf \nabla} \theta)^2 d^2 {\bf x},
\end{equation}
where $\rho_s$ is the coarse-grained spin-wave stiffness of the system
\cite{Kosterlitz73}.
The equation of motion (\ref{EQN:SPINDYN}) simplifies to
\begin{equation}
\label{EQN:ANGLEDYNAMICS}
\dot{\theta}_{\bf k} = -\rho_s \Gamma k^2 \theta_{\bf k} + \xi_{\bf k},
\end{equation}
where we only keep the component of the thermal noise locally orthogonal
to the order parameter $\vec{\phi}$, with
$\left< \xi_{\bf k}(t) \xi_{{\bf k}'}(t') \right>
= 2 \Gamma k_B T \delta_{{\bf k},-{\bf k}'} \delta(t-t')$.
[We will adsorb $\Gamma$  and $\rho_s$ into the time-scale, making $[t]$
dimensionally equivalent to $[l]^2$ for the rest of the paper, except
when we discuss quenches with general temperature history.]
We have ignored the $2 \pi$ periodicity of $\theta$ in equation
(\ref{EQN:ANGLEDYNAMICS}) because our system
has no vortices after the coarse-graining,
so that the phase can be taken to be everywhere continuous.

The solution to the equation of motion (\ref{EQN:ANGLEDYNAMICS}) is
\begin{equation}
\label{EQN:THETAT}
	\theta_{\bf k}(t) = \theta_{\bf k}(0)e^{-k^2 t}+
		\int_0^t d \tilde{t} e^{-k^2(t-\tilde{t})}
		\xi_{\bf k}(\tilde{t}),
\end{equation}
where the time $t \geq 0$ is measured from the time of the quench.
Since we start from an equilibrated state at or below $T_{\text KT}$,
the initial phases are determined by the spin-wave Hamiltonian
(\ref{EQN:SWHAM}).  The Fourier transformed phases are Gaussian
distributed with a probability distribution
\begin{equation}
\label{EQN:THETAINIT}
P[ \{ \theta_{\bf k}(0) \} ] \propto \exp \left\{ - \sum_k \frac{k^2}
{4 \pi \eta_i}
\theta_{\bf k}(0) \theta_{-{\bf k}}(0) \right\}.
\end{equation}
We use $\eta_i$ and $\eta_f$ to
describe the initial and final quench states, respectively, of our
system, with $\eta(T) = k_B T / 2 \pi \rho_s(T)$. They describe the
decay of equilibrium correlations and so are directly measurable in
experiments. The ``temperature'' $T$ always indicates the combination of
system parameters (temperature, pressure, composition, \ldots)
that determines $\rho_s$ and $\eta$.

The phase-phase correlations at general times after a quench at $t=0$
from a temperature $T_i$ to a temperature $T_f$, both at or below
$T_{\text{KT}}$ \cite{TKT}, are then straight-forward to calculate from
(\ref{EQN:THETAT}):
\begin{equation}
\label{EQN:CORRTHETA}
	\left< \theta_{\bf k}(t) \theta_{-{\bf k}}(t') \right> =
	\frac{2 \pi}{k^2} \left[ \eta_f e^{-k^2 |t'-t|}
	+ (\eta_i-\eta_f) e^{-k^2 (t+t')} \right].
\end{equation}
In particular, the initial correlations are given by
$\left< \theta_{\bf k}(0) \theta_{-{\bf k}}(0) \right> =
2 \pi \eta_i/ k^2$. The phase-difference correlations then follow,
\begin{eqnarray}
\label{EQN:THETA2}
	B(r,t,t') &\equiv&
	\left< [ 	\theta({\bf x},t)-
				\theta({\bf x}+{\bf r},t') ]^2 \right>
	\nonumber \\
	&=& \int \frac{d^2 {\bf k}}{(2 \pi)^2} \left[
	\left< \theta_{\bf k}(t) \theta_{-{\bf k}}(t) \right>
	+ \left< \theta_{\bf k}(t') \theta_{-{\bf k}}(t') \right>
	-2 \cos{({\bf k} \cdot {\bf r})}
	\left< \theta_{\bf k}(t) \theta_{-{\bf k}}(t') \right>
	\right] \nonumber \\
  &=& B_{EQ}(r,t,t') + B_{NEQ}(r,t,t').
\end{eqnarray}
$B_{EQ}$ and $B_{NEQ}$ are the equilibrium and non-equilibrium
correlations at the final temperature, given by the terms proportional
to $\eta_f$ and $\eta_i-\eta_f$, respectively, in (\ref{EQN:CORRTHETA}):
\begin{equation}
\label{EQN:BEQ}
	B_{EQ}(r,t,t') = \eta_f \left\{
\gamma+\ln{(r^2/4a_0^2)} +{\text E}_1[r^2/4(a_0^2+|t-t'|)] \right\},
\end{equation}
and
\begin{equation}
\label{EQN:BNEQ}
	B_{NEQ}(r,t,t') = (\eta_i-\eta_f) \left\{ {\gamma}
	+\ln{\left( \frac{r^2}{4 \sqrt{(a_0^2+2t)(a_0^2+2t')}}\right)}
		+{\text E}_1 [ r^2/4(a_0^2+t+t')]
		\right\},
\end{equation}
where
$\text{E}_1(x) \equiv  \int_x^\infty dy e^{-y}/y$
for $x>0$ and ${\gamma} \simeq 0.577$ is Euler's constant.
[We have used a soft ultra-violet cutoff, through a factor
$\exp(-a_0^2k^2)$ in the integrand of each ${\bf k}$-integral,
where $a_0$ is of order the lattice spacing.]
Both $B_{EQ}$ and $B_{NEQ}$ are manifestly symmetric
under interchange of $t$ and $t'$. $B_{EQ}$ only depends on the
magnitude of the time-difference $|t-t'|$, as expected for an
equilibrium
correlation.  Conversely, $B_{NEQ}$ has a scaling form at late times,
\begin{equation}
\label{EQN:SCALE}
B_{NEQ}(r,t,t') = F(r/\sqrt{t},r/\sqrt{t'}), \ \ \ \ \ \ t,t' \gg a_0^2,
\end{equation}
where $F(x)$ is a time-independent scaling function.

Order-parameter correlations follow directly from the
phase correlations since the phase variables are Gaussian distributed
at all times --- due to the linear evolution equation
(\ref{EQN:THETAT}), and the Gaussian nature of the noise and the
initial conditions. The general two-point two-time correlation function
is
\begin{eqnarray}
\label{EQN:CORR}
	C(r,t,t')
	&\equiv& \left< \vec{\phi}({\bf x},t) \cdot
		\vec{\phi}({\bf x}+{\bf r},t') \right>, \nonumber \\
	&=& \left< \cos{( [ \theta({\bf x},t)-
			    \theta({\bf x}+{\bf r},t') ])}
			 \right>, \nonumber \\
	&=& \exp \left\{- B(r,t,t') /2 \right\} \nonumber \\
	&=& C_{EQ}(r,|t-t'|) C_{NEQ}(r,t,t'),
\end{eqnarray}
where $C_{EQ}(r,|t-t'|) \equiv \exp \{ - B_{EQ}(r,t,t')/2 \}$ and
$C_{NEQ}(r,t,t') \equiv \exp \left\{ - B_{NEQ}(r,t,t') /2 \right\}$.
We see that $C(r,t,t')$ has a product scaling form, with a growing
length-scale $L \propto t^{1/2}$ characterizing the
non-equilibrium factor $C_{NEQ}$ through Eq. (\ref{EQN:SCALE}).
For quenches to $T_f>0$, the equilibrium factor $C_{EQ}$ has a
non-trivial form equal to the equilibrium correlation function of the
critical
point at the final temperature.  This is a generalization of the
scaling expected for a quench to within an ordered phase, dominated by a
$T=0$ fixed point, where $C_{EQ}(r) = \langle |\vec{\phi}| \rangle^2$.
We do not expect the product form seen in  Eq. (\ref{EQN:CORR}) to hold
in general $O(n)$ systems for
quenches to critical points; the product form holds for this  XY
system due to the Gaussian distribution of the phase variables.

It is interesting
to contrast these scaling results to the scaling prediction
$L \sim (t/\ln{t})^{1/2}$ \cite{Bray94a}, for quenches with free
vortices,
observed in simulations \cite{Yurke93} (note also \cite{other}).
Quenches from below $T_{KT}$ {\em are} consistent
with previous Energy-Scaling predictions \cite{Bray94a}, which are based
on the observed late-time defect structure for quenches to $T=0$.
For non-conserved quenches {\em without} topological defects, a growth
law $L(t) \sim t^{1/2}$ is predicted for systems which scale
\cite{Bray94a} -- in agreement with the results of this paper. For
quenches to $T>0$, the Energy-Scaling approach does not directly apply.
However, it is reassuring
that scaling and the same growth law is observed in the non-equilibrium
correlations $C_{NEQ}$.

We determine the asymptotic correlations by using the
asymptotics of $\text{E}_1(x)$:
\begin{equation}
\label{EQN:ASYMPT}
	\text{E}_1(x) \sim \left\{ \begin{array}{c}
			- {\gamma} - \ln{x}, \ \ \ \ \ \ x \ll 1, \\
			e^{-x}/x, \ \ \ \ \ \ \ \  \ \ \ x \gg 1.
			\end{array} \right.
\end{equation}
These determine the asymptotic equilibrium correlations,
\begin{equation}
\label{EQN:EQ}
C_{EQ}(r,0) \sim (r/a_0)^{-\eta_f}, \ \ \ \ \ r \gg a_0,
\end{equation}
which reproduces the standard result \cite{Kosterlitz73}.  The full
equal-time correlations after the quench have the asymptotic behavior
\begin{eqnarray}
\label{EQN:C}
C(r,t,t) &\simeq& \left\{ \begin{array}{c}
	 	(r/a_0)^{-\eta_i} (\sqrt{t}/a_0)^{\eta_i-\eta_f},
			\ \ \ \ \ \ \ \ \ \ \ r^2/t \gg 1, \\
			(r/a_0)^{-\eta_f}, \ \ \ \ \ \ \ \  \ \ \
		\ \ \ \ \ \ \ \ \ \ \ \ \  \ \ \ \ \ r^2/t \ll 1, \\
			\end{array} \right.
\end{eqnarray}
where $r, \sqrt{t} \gg a_0$.  They have the same spatial dependence as
the equilibrium correlations (with the final temperature determining the
correlations at short distances and the initial conditions determining
them at long distances), but have an additional amplification
factor at long scales.  For $\eta_i = \eta_f$ we recover
the equilibrium correlations (\ref{EQN:EQ}), as expected.

The autocorrelation function is given by the $r=0$ correlations with the
initial conditions. From equations (\ref{EQN:THETA2}) to
(\ref{EQN:ASYMPT}),
\begin{equation}
\label{EQN:AUTOCORR}
	A(t) \equiv C(0,t,0)
 	 \sim \left(\frac{t}{a_0^2}
	\right)^{- (\eta_i+\eta_f) /4},
\end{equation}
where again we take $t \gg a_0^2$.
Using $L \sim t^{1/2}$, and
$A(t) \sim L^{-\bar{\lambda}}$ \cite{Fisher88},
we determine the exponent $\bar{\lambda} = (\eta_i+\eta_f)/2$
describing the decay of autocorrelations after a quench.  Interestingly,
 $\bar{\lambda}$ does not change if the quench is reversed:
from $\eta_f$ to $\eta_i$.  For the decay  of equilibrium
autocorrelations, we set $\eta_f = \eta_f = \eta$ and find
$\bar{\lambda}_{EQ} = \eta$.

Bray {\em et al.} \cite{Bray91} have considered
quenches from long-range correlated initial conditions to
$T_f = 0$ ($\eta_f=0$). They predict that if initial long-range
correlations are described by $C(r,0) \sim r^{\sigma-d}$, and if
$\sigma < \sigma_c$, then $C(r,t) \sim (L/r)^{d-\sigma}$ for
$ r \gg L$ and $A(t) \sim t^{(\sigma-d)/4}$ for large $t$.
For quenches from a critical point,  $\sigma = 2 - \eta$.
Our results for quenches to $T=0$ agree  (with $d=2$):
from Eq. (\ref{EQN:C})
we have $C \sim (L/r)^{\eta_i}$ with $L \sim t^{1/2}$, while from
Eq. (\ref{EQN:AUTOCORR}) we have  $A(t) \sim t^{-\eta_i/4}$.
The initial correlations are relevant for quenches from all
temperatures at or below $T_{KT}$, so that
$\sigma_c \leq 2 - \eta(T_{KT}) = 7/4$.
Because $\bar{\lambda}_{SR}$, the
value for short-range correlated initial conditions, is never less than
$\bar{\lambda}$ for long-range correlated initial conditions
\cite{Bray91}, we have
$\bar{\lambda}_{SR} \geq \bar{\lambda}(T_{KT}) = 1/8$
 for a quench to $T_f=0$.
This lower bound is much smaller than the values measured by Pargellis
{\em et al.} \cite{Pargellis94}.  Theoretical treatments of quenches
{\em to} critical points \cite{Janssen89} have previously only
treated the case of a quench with uncorrelated initial conditions,
where $\bar{\lambda}$ is independent of the details of the initial
conditions. The methods of reference \cite{Bray91} can be extended to
general systems, in $d$-dimensions, with a quench to a general
critical point $T_f$ (characterized by an equilibrium correlation
function $C(r) \sim r^{-(d-2+\eta_f)}$), from a state with power-law
spatial correlations (in the same order-parameter field) decaying as
$r^{-(d-2+\eta_i)}$.  We find $\bar{\lambda} = d-2+(\eta_i+\eta_f)/2$
\cite{known}, provided that the long-range correlations in the initial
state are {\em relevant}, which requires that the exponent
$\bar{\lambda}_{SR}$ for a quench to $T_f$ from initial
conditions with {\em short-range} correlations satisfy
$\bar{\lambda}_{SR} < d + (\eta_f-\eta_i)/2$.
For quenches between two critical points in the 2D XY model,
our result $\bar{\lambda} = (\eta_i +\eta_f)/2$ agrees with the
general result for $d=2$, and depends on both the
intitial and the final temperatures of the quench.

Arbitrary temperature histories of the quench can be treated in a
similar
manner to our discussion so far.  We assume that the thermal bath has a
well-defined time-dependent temperature $T=T(t)$, i.e. that
microscopic time-scales are much faster than the phase-ordering
time-scales.  Then the renormalized spin-wave stiffness,
$\rho_s(T)$, and the equilibrium decay exponent, $\eta(T)$, will both be
time-dependent through their temperature dependence.
Eq. (\ref{EQN:SPINDYN}) to (\ref{EQN:ANGLEDYNAMICS}) will be
unchanged, while Eq. (\ref{EQN:THETAT}) will change to
\begin{equation}
\label{EQN:THETATT}
	\theta_{\bf k}(t) = \theta_{\bf k}(0)e^{-k^2 p(t)}+
		\int_0^t d \tilde{t} e^{-k^2 [p(t)-p(\tilde{t}) ]}
		\xi_{\bf k}(\tilde{t}),
\end{equation}
where $p(t) \equiv \int_0^t \rho_s(\tilde{t}) d \tilde{t}$.
This leads to phase correlations
\begin{equation}
\label{EQN:CORRTHETAT}
	\left< \theta_{\bf k}(t) \theta_{-{\bf k}}(t') \right> =
	\frac{2 \pi \eta_i}{k^2} e^{ - k^2 [ p(t) + p(t')] }
	+ 2 k_B \int_0^{\text{min}(t,t')} d \tilde{t}
	e^{- k^2 [ p(t) + p(t') - 2 p(\tilde{t}) ] } T(\tilde{t}).
\end{equation}
This can be used in Eqs. (\ref{EQN:THETA2}) and (\ref{EQN:CORR}) to
determine correlations under an arbitrary temperature history.  It is
straight-forward to show that temperature changes before a time
$t_M$ do not affect correlations with $t,t' \gg t_M$ or autocorrelations
$A(t)$ for $t \gg t_M$.  This is best illustrated
by considering autocorrelations
in the low-fugacity limit, where $\rho_s$ is temperature independent.
With some work, we have
$B(0,0,t) = (\eta_i/2) \ln{(t/a_0^2)} + \int_0^t d
\tilde{t} \eta(\tilde{t})/ (a_0^2 + 2t- 2 \tilde{t})$. [We have adsorbed
the constant $\rho_s$ into $\eta(t)$.]  If we restrict the
time-dependence
of $T$ to times before $t_M \ll t$, then $B(0,0,t) = \ln{(t/a_0^2)}
(\eta_i+\eta_f)/2 + O(t_M/t)$ and we recover our previous result.  The
same approach demonstrates that late time correlations
and autocorrelations are insensitive to the temperature history
before a time $t_M \ll t$.
Exact correlations for simple quench histories can easily be calculated,
and may be useful for comparing to experiments, but do not
appear qualitatively different than instantaneous quenches.

The ease of calculation of this model is fortuitous. In part this is
because, for vector systems, the gradient term in equation
(\ref{EQN:HAMILTONIAN}) dominates over the potential term
\cite{Bray94a}.
Since the potential term is not needed to set a core-scale for
singular defects (such as free vortices), the hard-spin limit may be
taken without complications.
The 2D XY model is special below $T_{KT}$ because the vortices are
tightly bound and only serve to renormalize the spin-wave stiffness
$\rho_s$ of the effective hard-spin Hamiltonian.  The resulting
Gaussian nature of the phase variables greatly simplifies the analysis.
[Note that the hard-spin limit for conserved ``model B''
dynamics \cite{Rutenberg94} leads to a much
more complicated evolution equation than (\ref{EQN:ANGLEDYNAMICS}),
and the phases will not be Gaussian distributed.]
Of course our results only apply for distances large with respect to the
vortex pair size in the two equilibrium phases at $T_i$ and $T_f$
\cite{pairs}.

The specially prepared nematic system of Pargellis {\em et al.}
\cite{Pargellis94}, developed to exhibit 2D XY behavior, should
exhibit the behavior described by this paper.
The experimental procedure will be easier since late-time correlations
and autocorrelations do not depend on the early stages of the quench.
[Experimental analysis will not be able to rely, however, on the
characteristic schlieren patterns of free vortices, which will
be absent.] Most other planar nematic liquid crystal systems,
while they exhibit an XY-like  growth law
when free vortices are present \cite{Bray94a}, will not have their
vortex-free quenches described by this paper.  This is
because the spin-waves of an unconstrained
nematic order parameter are not of the $O(2)$
variety -- the spins are not characterized by a single angle.

We have calculated correlations for quenches in the 2D XY model
between any two phases at or below the Kosterlitz-Thouless temperature.
The non-equilibrium part of the correlations scales
with a characteristic length-scale $L(t) \sim t^{1/2}$.  This growth
law differs by a logarithmic factor from the growth laws expected for
quenches involving free vortices. The autocorrelation decay exponent
depends on both the initial and the final state of the quench,
$\bar{\lambda} = (\eta_i+\eta_f)/2$. The asymptotic
autocorrelations and equal-time correlations do not depend on the early
temperature history of the quench, but only on the initial conditions
(through $\eta_i$) and on the temperature at late times
(through $\eta_f$).

We thank T. Blum and D. A. Huse for discussions,
and the Isaac Newton Institute for hospitality.


\begin{references}
\bibitem{TKT} In terms of the exponents describing the equilibrium
correlations, we consider quenches between any two states with
$\eta \leq 1/4$, the value at $T_{KT}$ \protect\cite{Kosterlitz73}.
\bibitem{pairs} At distances comparable to the vortex pair size
 the bound vortices will affect the correlations. In
particular, the Fourier transformed correlations $S(k,t)$
will have a weak power-law Porod tail $k^{-4}$, with amplitude
proportional to the vortex pair density for $k \rightarrow \infty$.
\end{references}
\end{document}